\documentclass{emulateapj}

\begin{document}

\title{Effects of Magnetic Field Strength and Orientation on Molecular Cloud Formation}

\author{Fabian Heitsch\altaffilmark{1}}
\author{James M. Stone\altaffilmark{2}}
\author{Lee W. Hartmann\altaffilmark{1}}
\altaffiltext{1}{Department of Astronomy, University of Michigan, 
                 Ann Arbor, MI 48109-1042, U.S.A}
\altaffiltext{2}{Department of Astrophysical Sciences, Princeton University, 
                 Princeton, NJ 08544, U.S.A.}
\lefthead{Heitsch et al.}
\righthead{Magnetized Molecular Cloud Formation}

\begin{abstract}
We present a set of numerical simulations addressing the effects of magnetic field
strength and orientation on the flow-driven formation of molecular clouds.
Fields perpendicular to the flows sweeping up the cloud can efficiently prevent
the formation of massive clouds but permit the build-up of cold, diffuse filaments.
Fields aligned with the flows lead to substantial clouds, whose degree 
of fragmentation and turbulence strongly depends on the background field strength. 
Adding a random field component leads to a ``selection effect'' for molecular cloud 
formation: high column densities are only reached at locations where the field 
component perpendicular to the flows is vanishing. Searching for signatures of colliding
flows should focus on the diffuse, warm gas, since the cold gas phase making up the
cloud will have lost the information about the original flow direction because the
magnetic fields redistribute the kinetic energy of the inflows. 
\end{abstract}

%%%%%%%%%%%%%%%%%%%%%%%%%%%%%%%%%%%%%%%%%%%%%%%%%%
%
%\section{Motivation}
%
%%%%%%%%%%%%%%%%%%%%%%%%%%%%%%%%%%%%%%%%%%%%%%%%%%
\section{Motivation}\label{s:motivation}

Evidence is accumulating that star formation follows rapidly upon molecular
cloud formation (e.g. \citealp{2001ApJ...562..852H} and \citealp{2007RMxAA..43..123B}
for the solar neighborhood; \citealp{2003ApJS..149..343E} for M33;
\citealp{2007ApJ...668.1064E} in the context of M51). This rapid onset
suggests that the clouds need to acquire high, non-linear density enhancements
during their formation, since massive, finite clouds are highly susceptible
to global gravitational collapse which could overwhelm small-scale fragmentation
necessary for (local) star formation \citep{2004ApJ...616..288B}. 
Thus to understand the initial conditions for star formation, we need to 
understand the formation of the parental clouds.
 
\citet{1999ApJ...527..285B} and \citet{2001ApJ...562..852H} proposed that the build-up
of clouds in large-scale, converging flows of diffuse atomic gas
could explain the crossing time problem, i.e. the observation that the stellar age 
spreads in a large number of local star forming regions are substantially smaller 
than the lateral crossing time \citep{2001ApJ...562..852H,2007RMxAA..43..123B}. 
In this picture, there need not be a causal connection between star formation events 
in the plane perpendicular to the large-scale flows (see also \citealp{2000ApJ...530..277E}). 
Rapid star formation is a necessary requirement for this scenario to work.

Numerical models of flow-driven cloud formation (we give only an early and the most
recent numerical work of each group, namely 
\citealp{1999A&A...351..309H} and \citealp{2008A&A...486L..43H};
\citealp{2000ApJ...532..980K} and \citealp{2008ApJ...687..303I};
\citealp{2005ApJ...633L.113H} and \citealp{2008ApJ...674..316H}; 
\citealp{2006ApJ...643..245V} and \citealp{2007ApJ...657..870V})
have identified the thermal and dynamical instabilities that are responsible for the 
rapid fragmentation of the  nascent cloud (see \citealp{2008ApJ...683..786H} 
for an assessment of the roles of the physical processes). Despite these promising successes, 
many questions about the physics at play remain unanswered, among one of the most pressing is  
the role of magnetic fields during the cloud formation process. 

The role of magnetic fields in the flow-driven cloud formation scenario has been largely
envisaged as one of ``guiding the flows'' to assemble the clouds, whether in
form of the Parker instability along galactic spiral arms 
\citep{1966ApJ...145..811P,1967ApJ...149..517P,1974A&A....33...73M},
in a generally turbulent interstellar medium (ISM) 
\citep{1995ApJ...455..536P,2001ApJ...562..852H}, 
or during the sweep-up of gas in spiral shocks \citep{2006ApJ...646..213K,2008MNRAS.383..497D}.
Based on the models of Passot et al., Hartmann et al. suggested that the field orientation
with respect to the flows selects the locations of cloud formation, namely that
clouds will only form if the fields are aligned with the flows. A perpendicular
field will reduce the compression of the post-shock gas and thus will limit the strong cooling and 
the thermal instability (TI, \citealp{1965ApJ...142..531F}) necessary for the rapid 
flow fragmentation and the build-up of high-density contrasts 
\citep{2004ApJ...616..288B,2008ApJ...674..316H,2008ApJ...683..786H}. 

Given sufficiently high strengths,
fields aligned with the inflows can suppress the dynamical instabilities responsible
for the generation of turbulence, namely the non-linear thin shell instability (NTSI, 
\citealp{1994ApJ...428..186V}; for a magnetic version see \citealp{2007ApJ...665..445H}) 
and the Kelvin-Helmholtz instability (KHI, e.g. \citealp{1961hhs..book.....C}, 
more recently \citealp{2008MNRAS.385.1494K}, and for numerical studies \citealp{2008ApJ...678..234P}).
Yet magnetic fields are intrinsically three-dimensional, and already two-dimensional
models by \citet{2008ApJ...687..303I} show that even for fields perpendicular to the inflow,
cold (albeit diffuse) clouds can form. 

Thus, three-dimensional models of flow-driven cloud formation including magnetic fields 
are needed. \citet{2008A&A...486L..43H} present a first approach to the problem, 
modeling the formation of a cloud in converging, perturbed flows, including fields and
self-gravity. Here, we focus on the 
effects of magnetic field strength and orientation on the early stages
of flow-driven cloud formation. We work in the ideal MHD limit (i.e. we do not
explicitly consider ambipolar drift or resistivity), and we do not include gravity in the models.

All our models start out with field strengths below equipartition with the kinetic energy
of the inflows. At a factor of $4.3$ below equipartition -- corresponding to an
absolute field strength of $5\mu$G at flow densities and velocities of $1$~cm$^{-3}$ 
and $16$~km~s$^{-1}$ -- , the fields already suppress the dynamical instabilities 
(and thus the generation of turbulence) leading to slab-like molecular clouds, while weaker 
fields -- at $2.5\mu$G, corresponding to a factor of $17$ below equipartition -- lead to clouds 
closely resembling the hydrodynamical case, albeit with more coherent filaments. 
Fields at $0.5\mu$G perpendicular to the inflows suppress the build-up of massive clouds in the collision
plane, while they lead to the formation of diffuse, cold filaments perpendicular
to the mean field, reminiscent of the cold HI clouds discussed by \citet{2003ApJ...586.1067H}.
A tangled field allows the assembly of substantial column densities
in regions where the lateral field component is small or vanishing.
Our results are consistent with the notion 
that magnetic fields select the location of cloud formation \citep{2001ApJ...562..852H}.

\section{Technical Details \& Parameters}

\subsection{Athena}
Calculations were performed with Athena \citep{2005JCoPh.205..509G,2008JCoPh.227.4123G}, 
an unsplit, second-order accurate Godunov scheme, using the corner transport upwind method
\citep{2008JCoPh.227.4123G} and a linearized Roe solver \citep{1981JCoPh..43..357R}. 
The divergence of the magnetic field is kept zero by using constrained transport
\citep{1988ApJ...332..659E}. Dissipative terms (viscosity, heat conduction and resistivity) 
are not explicitly included.
For a detailed description and test results, the reader is referred to 
\citet{2005JCoPh.205..509G,2008JCoPh.227.4123G} and \citet{2008ApJS..178..137S}.

We implemented heating and cooling as an additional energy source term at 2nd order in time.
A tabulated cooling function provides the energy change rate as function of density and temperature
at each grid cell. We decided to keep the iterative approach we had used in our earlier studies
of cloud formation \citep{2005ApJ...633L.113H,2006ApJ...648.1052H,2008ApJ...674..316H}, 
with a slight modification. Instead of advancing the fluid evolution at the usual time step
given by the Courant-Friedrichs-Levy (CFL) condition and subcycling on the energy equation in case
the cooling timescale is shorter than the CFL timestep, we lower the CFL timestep according to 
\begin{equation}
  \Delta t = \Delta t_{CFL} \min(1,(\tau_c/\Delta t_{CFL})^p),
\end{equation}
with $0\leq p\leq 1$. For increasing $p$, small cooling timesteps will control the overall CFL
timestep. The earlier version (see references above) would be equivalent
to $p\equiv 0$. Yet this choice can lead to inconsistencies in the hydrodynamical evolution 
once the cooling timesteps get substantially shorter than the fluid timesteps, leading to a 
numerical overemphasis of the acoustic mode of the TI, since regions can cool substantially 
without accounting for the resulting pressure drop in the dynamics. While these inconsistencies 
may not affect the overall results, they turn out to affect the stability of the solution. 
For the models presented here, $0.5<p<0.7$ yields a stable and accurate solution. 
While the iterative approach is more time-consuming than a direct integration as suggested by 
e.g. \citet{2007ApJ...657..870V}, it allows us to take into account the increasing thermal 
timescale of a fluid parcel cooling down towards the thermal equilibrium line 
(see e.g. Fig~3 of \citealp{2008ApJ...683..786H}). This effectively keeps 
a fraction of $40$--$50$\% of the gas mass in the thermally unstable warm neutral medium (WNM), 
consistent with observed fractions (\citealp{2003ApJ...586.1067H}; also \S\ref{ss:thermals}).

\subsection{Setup and Parameters}
The initial conditions and flow parameter are similar to those used in our earlier
work \citep{2008ApJ...674..316H}.
Two identical uniform flows collide head-on at an interface whose position is perturbed
using an amplitude that is chosen randomly, but with the constraint that only the long
wavelengths (down to a quarter of the box size) are non-zero. The maximum perturbation
amplitude measures $10$\% of the box size. The flows are otherwise in thermal equilibrium
and do not have substructure. Thus we can test the most unfavorable conditions for 
substructure and turbulence generation in the resulting clouds. 

The simulation domain has a resolution of  $256^3$, and 
the linear box size measures $44$~pc in all cases, resulting in a nominal resolution of $0.18$~pc. 
The boundaries in $y$ (and $z$) are periodic, while those in $x$ are set to a constant inflow speed 
of $v_0=16$~km~s$^{-1}$, a density of $n_0=1$~cm$^{-3}$, a temperature of $T_0=8\times 10^3$~K and a 
magnetic field (if applicable). The adiabatic exponent $\gamma=5/3$.
The flow speed $v_0$ is consistent with shock speeds along streamlines passing through Galactic
spiral shocks at the solar circle \citep{1972ApJ...173..557S}.
 Our thermal pressure is a factor $2-3$ higher than the pressures estimated for the CNM/WNM 
\citep{2001ApJS..137..297J,2003ApJ...586.1067H}, increasing the ratio of thermal over 
magnetic pressure $\beta_{th}$ above observed values (\citealp{2005ApJ...624..773H}, hereafter HT05). 
Yet the measure relevant for the dynamical importance of magnetic fields is the ratio of the 
{\em ram} pressure over the magnetic pressure,
$\beta_{ram}$, since the instabilities responsible for turbulence generation and fragmentation
depend on the ratio of the flow velocities over the Alfv\'{e}n speed
(\citealp{1961hhs..book.....C}; \citealp{2007ApJ...665..445H}). In the diffuse, local 
ISM, kinetic and magnetic energy are in approximate equipartition ($\beta_{ram}\approx 1$; HT05), 
while in regions of ordered, large-scale flows -- such as Galactic spiral shocks or expanding
supernova shells -- with $v_0\gtrsim 10$~km~s$^{-1}$, $\beta_{ram}>1$ should be expected. 

There are four classes of models: hydrodynamical (series H), field aligned with flow (series X),
field perpendicular to the flow (series Y), and (series XR) a uniform field component aligned with 
the flow plus a random field component of similar size, consistent with (although a little smaller than) 
observed magnetic field 
strength estimates (e.g. \citealp{1996ASPC...97..457H}; \citealp{2004Ap&SS.289..293B};
\citealp{2006ChJAS...6b.211H}). 
In the latter series, we do not perturb the 
collision interface but rely on the tangled field component to trigger the dynamical instabilities.  
Table~\ref{t:param} summarizes the model parameters. Self-gravity is not included in the models.

To initialize the random field component, we set the amplitudes and phases of e.g. the $x$-component
of the (edge-centered) vector potential to 
\begin{equation}
  A_x(x,y,z) = \sum_{i,j,k=1}^{max} |k|^{-p}\sin(k_xx+k_yy+k_zz+\phi_{i,j,k}^x),\label{e:vecpot}
\end{equation}
where $|k|\equiv k_x^2+k_y^2+k_z^2$ and e.g. $k_x\equiv 2\pi i/L_x$ with the box length $L_x$. 
We set $p\equiv 4$, mimicking a (steep) turbulent energy spectrum as observed in detailed
numerical simulations of magneto-hydrodynamic turbulence (e.g. \citealp{2003MNRAS.345..325C}).
The wavenumbers $k_{x,y,z}$ are chosen such that $1\leq |k| \leq 4$, i.e. all
combinations of $(i,j,k)$ in the sum over $k$-space are used that satisfy the constraint on $|k|$. 
The phases $\phi_{i,j,k}^x$ in $k$-space are chosen from a uniform random distribution. Each vector
potential component $A_{x,y,z}$ requires a separate phase array $\phi^{x,y,z}$. 

This formulation in real space instead of in Fourier space (see e.g. 
\citealp{1998PhRvL..80.2754M}; \citealp{1998ApJ...508L..99S}; \citealp{2008ApJ...682L..97L} for
velocity fields) allows us to easily regenerate 
the vector potential (and the field) at the inflow boundaries by
\begin{equation}
  A_x(\pm L_x/2,y,z,t) = A_x(\pm (L_x/2+v_0t),y,z),\label{e:bc}
\end{equation}
where the negative value refers to the lower $x$-boundary, and the positive to the upper one. 
The face-centered fields are then computed from the vector potential by 
$\mathbf{B}=\nabla\times\mathbf{A}$. 

The choice of the wave-number range $1 \leq |k| \leq 4$ does not constitute a restriction
in terms of generality of our simulations, since the energy distribution over spatial scales
is determined by the (steep) power law index $p$. This is fortunate in a sense, since the 
generation of the boundary conditions (eq.~[\ref{e:bc}]) would consume substantially more time
if we had to sum over all available $|k|$ in equation~(\ref{e:vecpot}).

\begin{deluxetable}{lcccccc}
\tablewidth{0pt}
\tablecaption{Model Parameters\label{t:param}}
\tablehead{\colhead{Name}&\colhead{$B_{x0}$ [$\mu$G]}
                         &\colhead{$B_{y0}$ [$\mu$G]}
                         &\colhead{$B_{rms}$ [$\mu$G]}
                         &\colhead{$\beta_{th}$}
                         &\colhead{$\beta_{ram}$}}
\startdata
H      & $0.0$ & $0.0$ & $0.0$ & $\infty$ & $\infty$ \\ % 3Ha
X25    & $2.5$ & $0.0$ & $0.0$ & $4.5$    & $17$     \\ % 3Xa1
X50    & $5.0$ & $0.0$ & $0.0$ & $1.1$    & $4.3$    \\ % 3Xa0
Y05    & $0.0$ & $0.5$ & $0.0$ & $110$    & $430$    \\ % 3Ya2
XR25   & $2.5$ & $0.0$ & $2.5$ & $2.2$    & $8.5$       % 3XRa1
\enddata
\tablecomments{1st column: model name, 
2nd: magnetic field strength $B_x$; 3rd: $B_y$, 4th: random field $B_{rms}$, 
5th: thermal plasma $\beta$, 6th: ram plasma $\beta$.}
\end{deluxetable}

\subsection{Physical Interpretation of the Initial Conditions}

Obviously, our initial conditions are somewhat idealized, e.g. 
generally, the flows would be expected to have substructure, 
the flows might not be expected to collide always head-on, and the
magnetic fields will have parallel and perpendicular components with respect
to the inflows. Yet the initial conditions can be seen as idealized versions 
of different physical environments.

The case of uniform fields aligned with the inflows (models X25, X50) 
could be identified with the sweep-up of material by an expanding supernova 
shell along an ordered background field, or with the collision of two expanding 
shells in such a field. The initial field strength of $5\mu$G (model X50) 
is close to the local median (total) field strength in the CNM (e.g. 
HT05; \citealp{2005ASPC..343...64T}). 
Using Nakano \& Nakamura's (\citeyear{1978PASJ...30..671N})
expression for the critical surface density $N_c\equiv B/\sqrt{4\pi^2 G}$ 
above which gravitational collapse is possible under flux-freezing conditions, 
the swept-up  clouds would reach approximately $0.5N_c$ after $12$~Myr, while model
X25 ($B_{x0}=2.5\mu$G) would be marginally critical at the same time. We defer
the discussion of the mass-to-flux ratio in the clouds to a subsequent paper
including gravity. 

An ordered field aligned with the flows plus a large-scale random
component of similar amplitude (model XR25) introduces a large-scale shear and 
might be considered a general situation for sweep-up of gas in spiral shocks, 
while the perpendicular field case (Y05) would address the (probably common) 
situation of an oblique field whose lateral component is amplified by flow compressions.

We emphasize that while we attempt to address the extreme situations
of field orientations, the finite size of our simulation domains cannot fully capture the 
effects of the magnetic field's boundary conditions. These will be set on larger 
scales than our local simulations can cover. In that sense, our results should be 
viewed as providing insight into magnetized cloud formation under idealized conditions 
rather than under physically realistic ones.

\subsection{A Comment on Resolution}

We decided to keep the resolution of our models constant, foregoing a resolution
study in favor of a parameter study. Resolution effects have been discussed 
by \citet{2007A&A...465..431H}. In addition, we have performed a systematic
resolution study for two-dimensional cloud formation models (unpublished -- the
models are similar to the ones discussed by 
\citet{2005ApJ...633L.113H,2006ApJ...648.1052H}), covering a factor of $32$ in
spatial resolution (from $256^2$ to $8192^2$ cells). As has been pointed out,
the critical length scale to resolve is the cooling 
length of the thermal instability. If not resolved, the thermal instability will 
be partially suppressed. At parameters of the WNM, the cooling
length is on the order of a parsec, while for the cold neutral medium (CNM), 
it drops to a fraction of a parsec. Thus, while more substructures should form with
increasing resolution, we expect our models to follow the general evolution
of the thermal and dynamical instabilities sufficiently accurately for our purposes.

\section{Results}

\subsection{Morphologies}

\begin{figure*}
  \begin{center}
  \includegraphics[width=0.7\textwidth]{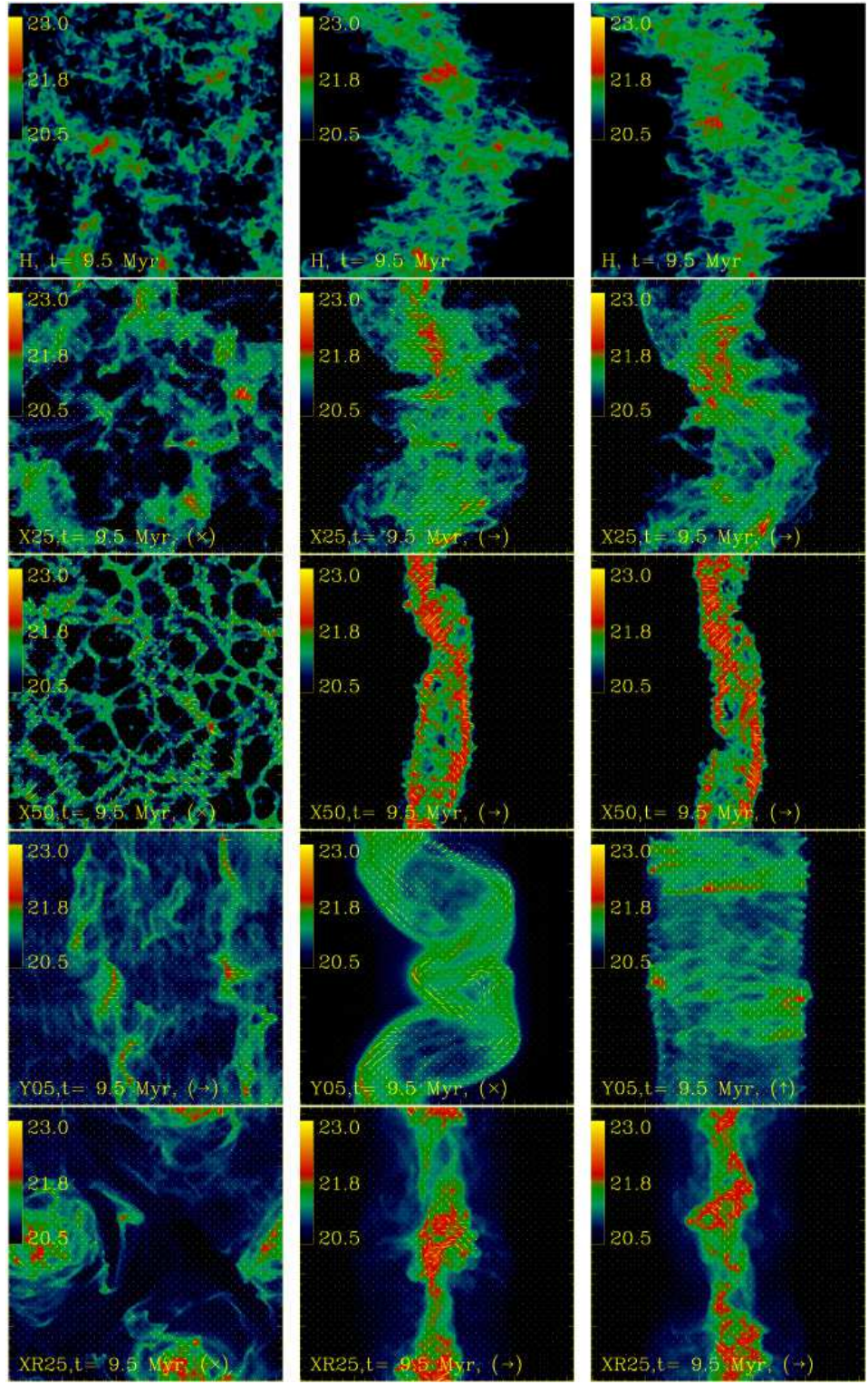}
  \end{center}
  \caption{\label{f:polmap}Logarithmic column density projections (in cm$^{-2}$) 
           along the three 
           grid axes ({\em left}: along inflow, {\em center} and {\em right}:
           perpendicular to the inflow) for models H, X25, X50, Y05 and XR25 as indicated,
           at $t=9.5$~Myr after flow collision. The mean field direction is denoted by 
           the symbols in the panel labels ($\rightarrow$,$\uparrow$,$\times$).}
\end{figure*}

Figure~\ref{f:polmap} summarizes the morphological effects of magnetic fields during
the build-up of a cloud. From top to bottom, it shows logarithmic column density maps of 
the hydrodynamical model H, and the four MHD models X25 through XR25. The three columns stand 
for projections along each coordinate axis, namely along the inflow ($x$-axis, {\em left}), 
and perpendicular to the inflow (along $y$ and $z$-axes, {\em center} and {\em right}). The 
maps of the MHD-models show polarization vectors which have been determined by integrating 
the density-weighted Stokes $Q$ and $U$ parameters along the respective line-of-sight 
(see \citealp{1996ASPC...97..486Z}; \citealp{2001ApJ...561..800H}). 

\subsubsection{Field parallel to inflow}

Model H shows the strong fragmentation due to thermal and dynamical instabilities
similar to the models discussed by \citet{2008ApJ...674..316H}. 
Specifically, the large-scale initial perturbation
triggers the NTSI, due to whose rapid growth some of the dense material has already reached
the inflow boundaries. Viewed along the inflow (top left panel in Fig.~\ref{f:polmap}), the
cold dense fragments appear clumpy rather than filamentary.

Introducing a magnetic field {\em aligned} with the flow (model X25, second row)
suppresses fragmentation compared to model H.
The face-on view ({\em left}) shows several large-scale coherent filaments with
denser cores. The edge-on views ({\em center} and {\em right}) 
demonstrate that the magnetic field is not dynamically
dominant. The polarization vectors are aligned with local structures.

Increasing the magnetic field (model X50, third row from top) 
seems to suppress much of the fragmentation. Specifically, the NTSI is only very weakly 
(if at all) present, since the magnetic field is
strong enough to suppress the lateral momentum transport necessary for triggering the NTSI
\citep{2007ApJ...665..445H}. Nonetheless, the flows still fragment, albeit into a tight 
network of filaments
(X50, left) instead of a few large, more clumpy and fuzzy structures 
(models H, X25). The suppression of the NTSI leads to the formation of a more or less coherent 
filament in the lateral projection (center and right column for model X50). Local shear modes 
lead to strong distortions of the field from its initial alignment with the inflow, as 
indicated by the polarization vectors which mostly trace out the mean background field.

\subsubsection{Field perpendicular to inflow}

The introduction of a field {\em perpendicular} to the inflow changes the
morphology completely (4th row of Fig.~\ref{f:polmap}, model Y05), despite the by a 
factor of $10$ 
weaker field (see Table~\ref{t:param}). The perpendicular field breaks the symmetry in the plane of
the flow collision, leading to filaments perpendicular to the mean field direction (note that
the mean field in the left panel of the 4th row of Fig.~\ref{f:polmap} is oriented horizontally). 
These filaments form due to motions along the field lines, but perpendicular to the incoming flows
(see also \citealp{2007ApJ...665..445H} and \citealp{2008ApJ...687..303I} 
for two-dimensional models). The magnetic field suppresses 
one degree of freedom in the gas motions, also leading to lower column density contrasts
than in models H and X50. The two lateral views of model Y05 exhibit another effect of 
the perpendicular field. Seen along the mean field direction, a large scale NTSI-driven mode 
is discernible, while
the projection perpendicular to the inflow and to the mean field (Y05 right) just shows a 
slab (albeit with substructure). In the former, the field lines are just shuffled around and
contribute to the dynamics only via the pressure term in the Lorentz force, thus lowering
the column densities and broadening the slab (compare to center panel of model X50).
In the latter, the tension term of the Lorentz force prevents the growth of the NTSI.
This is evidence for the presence of interchange modes in the NTSI, similar to e.g. 
the Rayleigh-Taylor instability \citep{2007ApJ...671.1726S}.
Still, material is free to move along the field lines (and thus perpendicular to the inflows), 
leading to the formation of the filaments {\em parallel} to the inflows. 

\begin{figure}
  \includegraphics[width=\columnwidth]{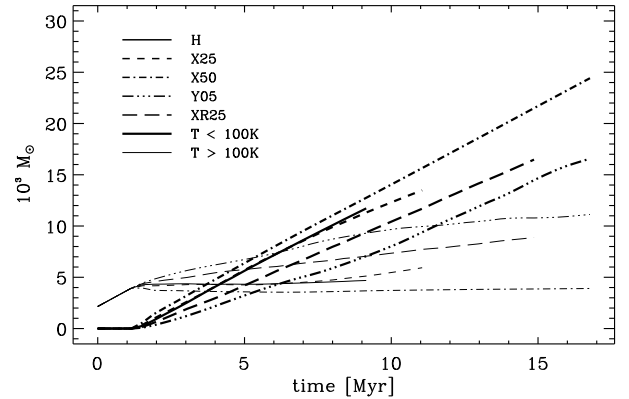}
  \caption{\label{f:masses}Total masses against time, below and above $T=100$~K. 
           The perpendicular field
           reduces the compression and thus lowers the cold gas mass, while the 
           field aligned with the flow leads to higher compressibility.}
\end{figure}

The magnetic field perpendicular to the inflow resists compression, leading
to a suppression of the thermal instability, which is also mirrored in the total mass budget
of all models (Fig.~\ref{f:masses}). Model X50 has the highest fraction of cold gas, due to 
the strong guide field leading to a strong compression of the gas, while model Y05 shows the
smallest cold mass fraction, because the lateral field resists compression by the flows, and thus
reduces the cooling rates. 

\subsubsection{Tangled field}

The bottom row of Figure~\ref{f:polmap} shows the maps for model XR25, which starts out
with a uniform field aligned with the flow at $2.5\mu$G and a random field component of 
equal magnitude. Although the (varying) lateral field components contain $10$ times
as much energy as the perpendicular field in model Y05, the fields do not suppress
the formation of clouds with column densities in excess of $10^{22}$~cm$^{-2}$; a tangled
field is substantially less efficient in preventing compression. Since there are regions
where  the field will be aligned with the flow, it leads 
to a selection effect in the sense that the clouds form at positions where
the lateral random components of the fields are weakest over time and/or where bends
in the fields determine the position of cloud formation (see \citealp{2001ApJ...562..852H}).
The resulting clouds 
are more isolated, with larger voids between them (bottom left panel of Fig.~\ref{f:polmap}).
The side view (bottom center and right) exhibits a diffuse halo of thermally unstable gas, 
material which is caught in the tangled field between the bounding shocks and the dense cold
gas. 

\subsection{Dynamics}

\begin{figure}
  \includegraphics[width=\columnwidth]{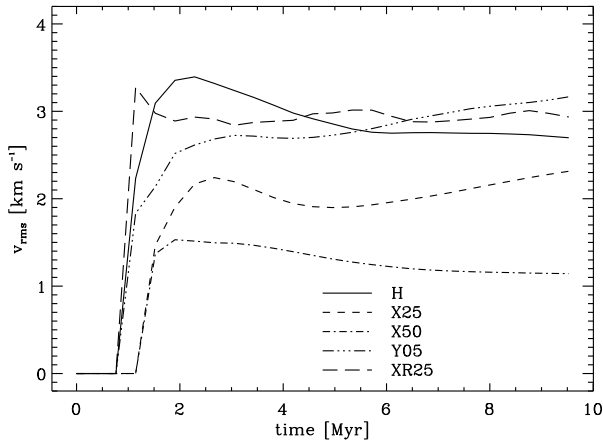}
  \caption{\label{f:vrms}Density-weighted $rms$ velocity dispersion against time for
           all models. Fields parallel to the inflows seem to suppress turbulence in the 
           cold gas.}
\end{figure}

\begin{figure*}
  \includegraphics[width=\textwidth]{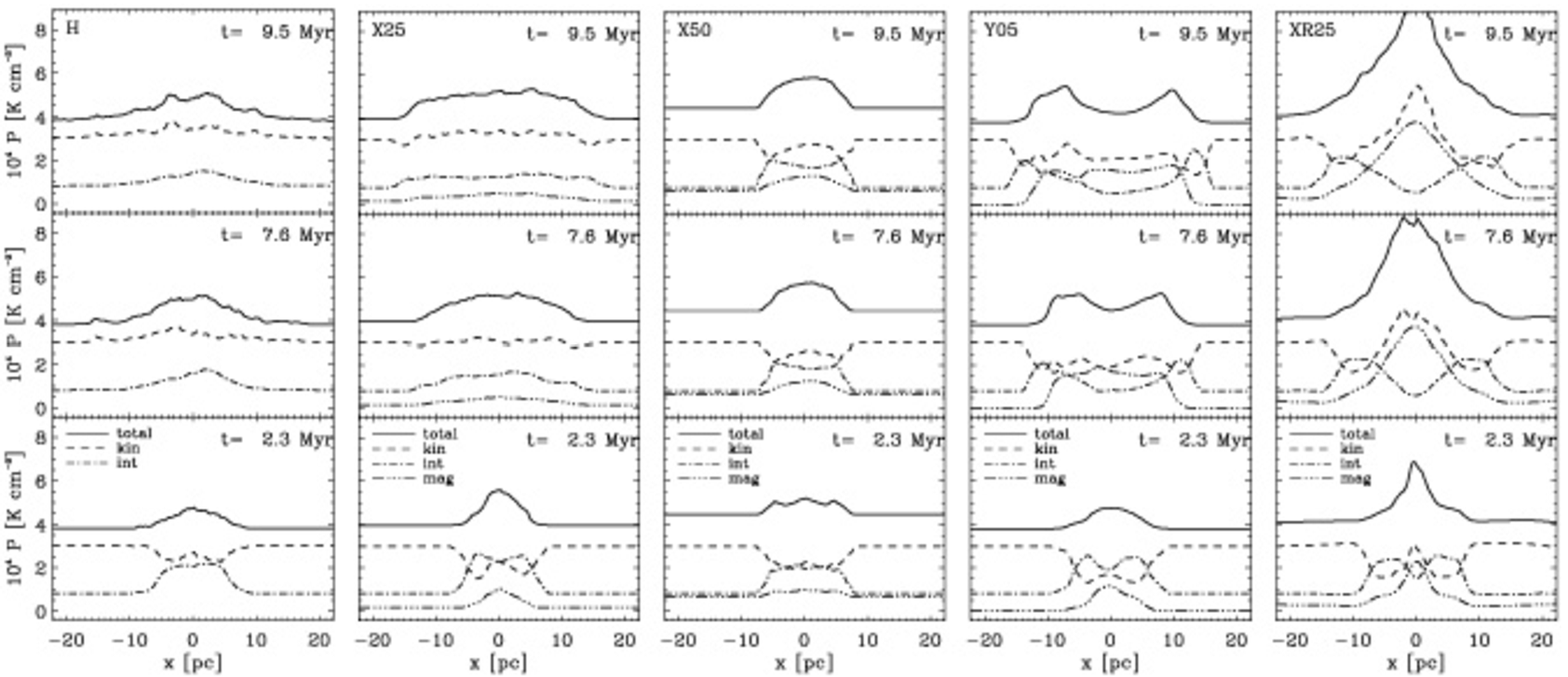}
  \caption{\label{f:prssprof}Pressure profiles along the inflow direction $(x)$, averaged over
           the perpendicular directions $(y,z)$, for three times as indicated, and for all models.
           Shown are total (solid line), kinetic (dashed),
           internal (dash-dot) and -- if applicable -- magnetic (dash-3-dot) pressures. Note 
           that we show the pressures, not the logarithm of the pressures.}
\end{figure*}

The cold mass fractions of models H and X25 (Fig.~\ref{f:masses}) are slightly lower than  that
of X50, indicating that the developing turbulence due to the flow fragmentation is also broadening
the slab. While this notion is already suggested by Figure~\ref{f:polmap}, it is confirmed by comparing
the $rms$ velocity dispersion in the cold ($T<300$~K) gas (Fig.~\ref{f:vrms}), and it can also be 
gleaned from a more detailed look at the laterally averaged pressure profiles (Fig.~\ref{f:prssprof}). 

Shown is a time sequence of the pressure profiles along the $x$-axis (i.e. along the inflows) for 
all models. In the absence of gravity, the slabs are all overpressured by the ram pressure of
the colliding flows (solid lines). At early times, all five models show a drop in kinetic pressure
and an increase in thermal pressure in the collision region. The flows have not fully fragmented
yet, and the cooling is not in full strength yet because of the still low densities. With evolving
time, the thermal pressure peak for model H drops due to increasing cooling. 

\begin{figure*}
  \includegraphics[width=\textwidth]{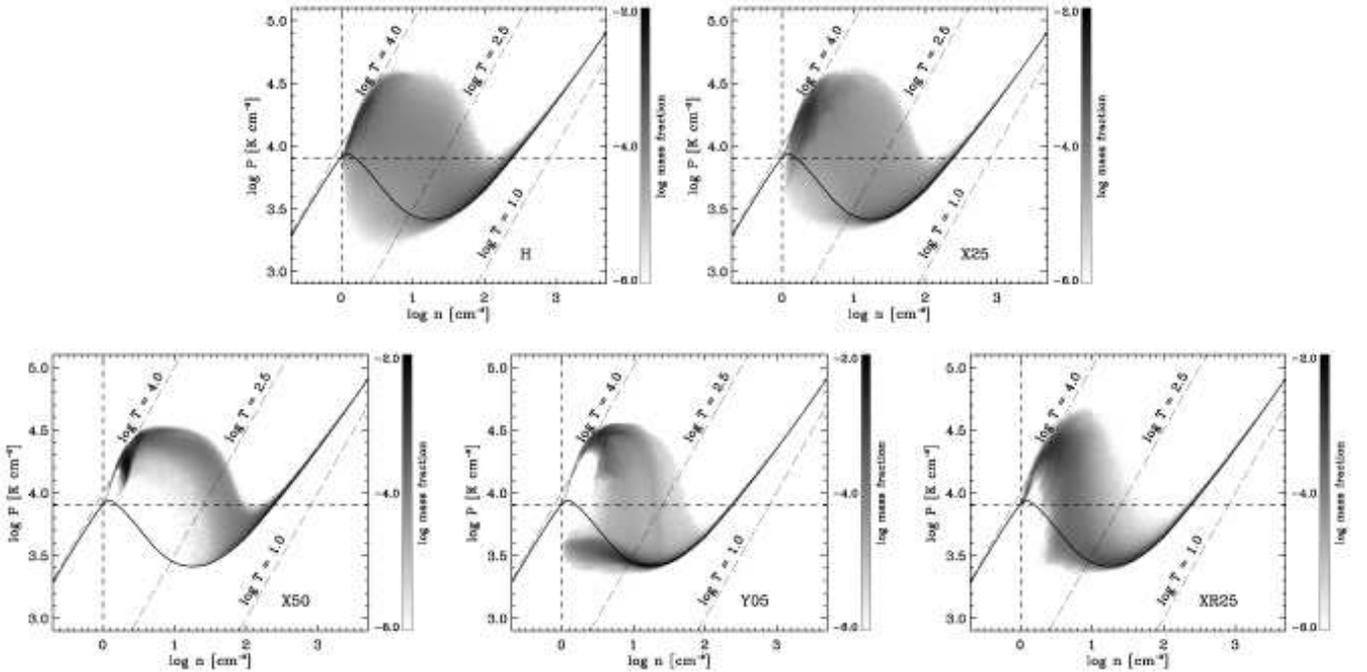}
  \caption{\label{f:prssdens}Greyscale-coded mass fraction of thermal pressure against volume density 
           for all models as indicated in the plots, at $t=9.5$~Myr. The solid line indicates the
           thermal equilibrium curve, while the vertical and horizontal dashed lines denote the 
           initial conditions in density and pressure. Diagonal dashed lines stand for isotherms
           at $T=10^4$, $300$ and $10$~K as indicated.}
\end{figure*}

This is markedly different for model X50, where the thermal pressure continues to be enhanced
by a factor of more than 2 above that of the inflow. Also, the kinetic pressure
drops, reaching an approximate equipartition with the thermal pressure. This is due
to the strong magnetic guide field, which ``splits'' the cloud into a network of dense filaments 
with low-density, high-temperature voids in between (see top panel of 2nd column of 
Fig.~\ref{f:polmap}, model X50). Figure~\ref{f:prssdens} offers a different view of the same
phenomenon, showing the pressure-density distributions for all four models, at $t=9.5$~Myr. 

The high-temperature voids of model X50 show up at $\log n \approx 0.5$ and $\log P \approx 4.3$, 
while the high-density filaments sit all on the stable low-temperature branch of the thermal
equilibrium curve at $T\approx 40$~K. Note that a substantial amount of the gas mass is actually
thermally over-pressured, in contrast to model H. 

Reducing the field aligned with the flow (model X25) leads to pressure profiles 
(Fig.~\ref{f:prssprof}) and thermal states (Fig.~\ref{f:prssdens}) similar to the hydrodynamical 
model H. In other words, while the field in model X25 is non-negligible in the sense that its 
presence still makes a morphological difference (see Fig.~\ref{f:polmap}), it does not noticeably 
affect the overall dynamics of the cloud.

The field is obviously dynamically important in model Y05. Because of the strong flow 
compression perpendicular to the field lines, the magnetic pressure takes over the role of the
thermal pressure, which leads to a substantial amount of thermally underpressured gas 
in model Y05 (Fig.~\ref{f:prssprof} and \ref{f:prssdens}). There is only a small amount of 
material at high ($\log n > 2$) densities.

Introducing the tangled field component on top of a field aligned with the inflows leads to an
over-pressurization of the slab by a factor of more than $2$ (Fig.~\ref{f:prssprof} right, model XR25).
This is mainly due to a combined increase in magnetic and kinetic pressure, i.e. the tangled field
leads to more turbulence than all other field geometries. The increase in kinetic pressure cannot
be solely due to enhanced densities -- model X50 should show a similar increase then. 
Although the tangled field in the diffuse gas phase is not force-free, it does not contribute
perceptibly to turbulent motions in the inflows, as can be seen by comparing the kinetic pressure
levels in the inflows between models XR25 and e.g. X50. Also, the kinetic pressure of model
XR25 does not increase when moving closer towards the midplane, until one enters the post-shock region.
The thermal pressure peaks in the diffuse
envelopes due to warm gas being unable to cool down (see bottom right panel of Fig.~\ref{f:prssdens}),
but it drops back to the ambient {\em thermal} pressure at the cloud midplane ($x=0$). Obviously,
the averaged thermal pressure alone is not a very accurate indicator of the cloud's physical state. 

\begin{figure*}
  \includegraphics[width=\textwidth]{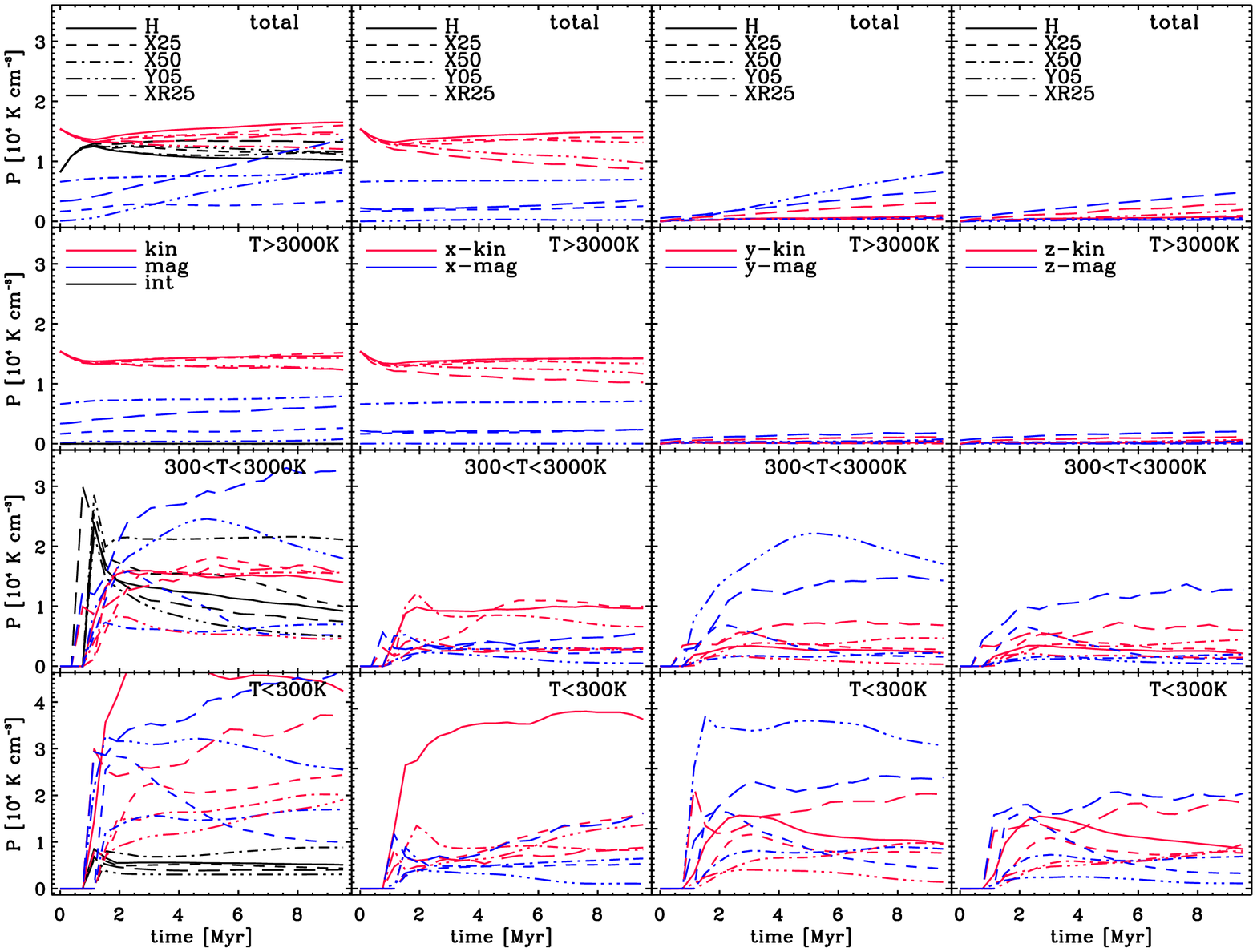}
  \caption{\label{f:prsstime}Pressures against time for all models, sorted according to temperature
          regimes (rows; total, $T>3000$K, $300<T<3000$K and $T<300$K) and 
          components (columns; total, $x$, $y$ and $z$). Black lines indicate thermal pressure, red
          lines kinetic and blue lines magnetic pressure. Note that these are again linear pressures in
          units of $10^4$~K~cm$^{-3}$.}
\end{figure*}

Figure~\ref{f:prsstime} summarizes the pressure evolution over time for all models. The pressures
are sorted according to temperature regimes (rows; total, stable WNM, unstable WNM, and CNM) 
and vector components (columns; total, $x$ parallel to
inflow, $y$ perpendicular to inflow, and $z$ again perpendicular to inflow). 
The total pressures ({\em top} row) stay constant with time except for the magnetic components of
models XR25 and Y05 due to the compression of the transverse field component (see $y$ column), 
and the $y$-component of the kinetic energy of model XR25. 
Not surprisingly, the stable warm neutral phase does not show any strong pressure variations with
time. In the thermally unstable regime, $300<T<3000$~K, the
models with transverse field components (XR25, Y05) show an increase in the $y$ and $z$ component
of the magnetic pressure. A large fraction of the magnetic energy in model XR25 is thus stored in
the diffuse cloud halo (see Fig.~\ref{f:polmap}). 
The magnetic fields also lead to a ``redistribution'' of kinetic energy between the vector components:
the bulk of the kinetic energy of the cold gas in model H is found in the $x$-component, parallel
to the inflow, indicating that the dynamics are dominated by the NTSI. The magnetic field models
have at least as much kinetic energy in the transverse ($y$ and $z$) components. For model
XR25, the transverse components even dominate. The substantial overpressure at the midplane
of model XR25 (see Fig.~\ref{f:prssprof}) arises mainly from the transverse components of the
magnetic and kinetic energy in the cold gas. While for $T<300$~K, the magnetic energies are larger
than the kinetic ones, the kinetic energy content dominates the total energy budget (top left
panel of Fig.~\ref{f:prsstime}). 

\begin{figure}
  \includegraphics[width=\columnwidth]{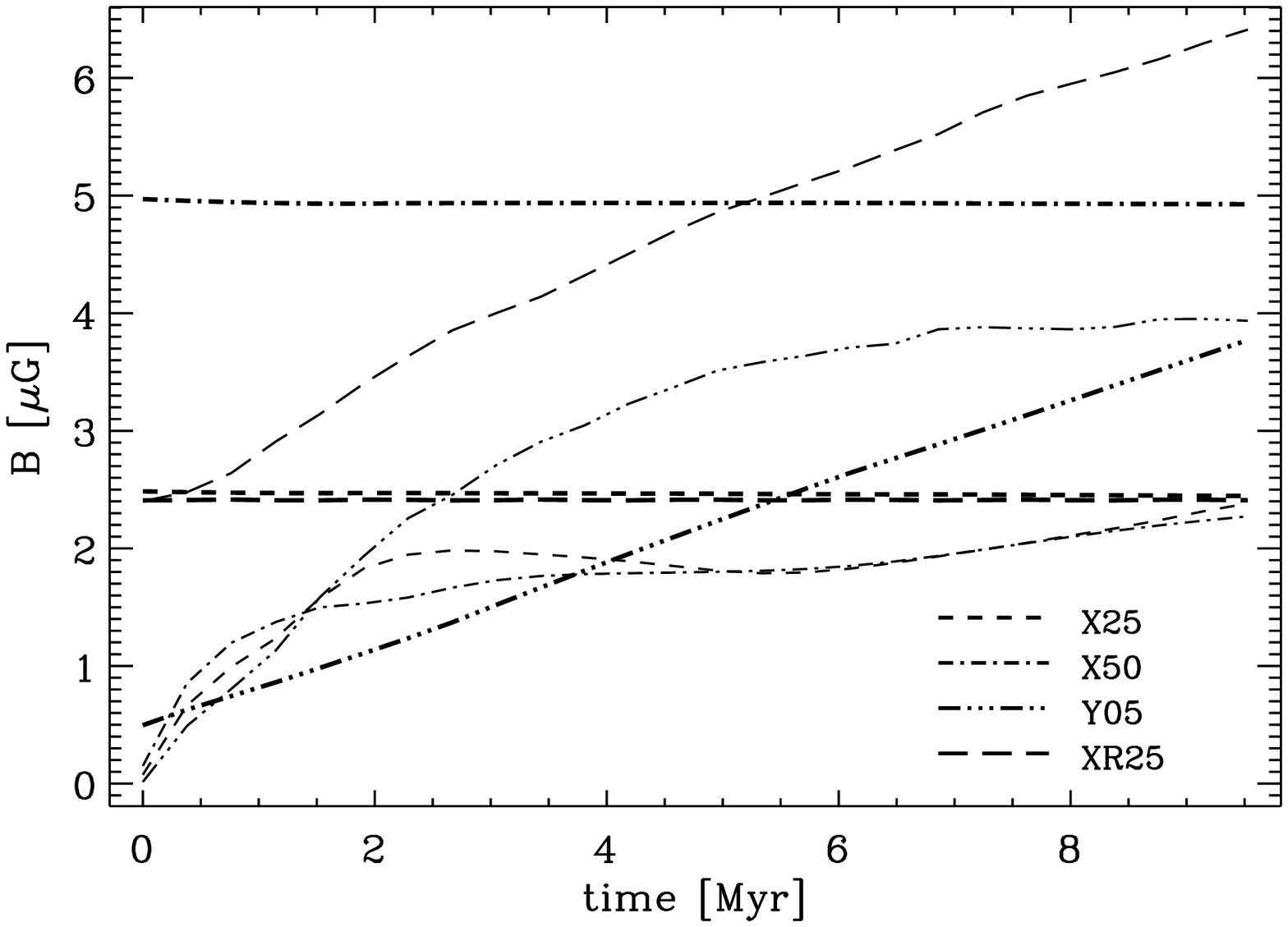}
  \caption{\label{f:fieldtime}Ordered (thick lines) and turbulent (thin lines) magnetic
           field components against time for models X25, X50, Y05 and XR25. For models
           X25 and XR25, the ordered and turbulent components are approximately equal.}
\end{figure}

Figure~\ref{f:fieldtime} offers a simpler view of the field amplification depending on mean field
direction. It shows the ordered (mean) field component and the random (turbulent) component 
against the time. Only the random components are growing with time, indicating that the fields
are mainly amplified by fieldline stretching. Only models Y05 and XR25 have field components
initially perpendicular to the inflows; these are growing linearly with time due to the compression
by the flows. The random (or turbulent) field components of models X25 and X50 evolve in a similar
way, indicating that a similar amount of energy is stored in the random component. For models
X25 and Y05, the ordered and random components reach similar levels (see \S\ref{ss:components}).

\subsection{Turbulence and Thermal States}\label{ss:thermals}

\begin{figure}
  \includegraphics[width=\columnwidth]{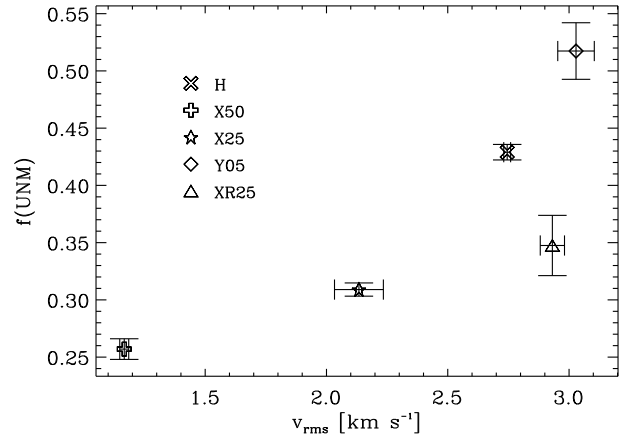}
  \caption{\label{f:vrmsunm}Mass fraction of thermally unstable gas ($300<T<3000$~K) 
           with respect to all WNM ($T>300$~K) against the velocity
           dispersion in the CNM (Fig.~\ref{f:vrms}), averaged between
           $7.5$ and $9.5$~Myr. The error bars show errors on the mean.}
\end{figure}

The ratio of gas mass in the thermally unstable regime to the mass in the WNM 
can be compared to observational constraints. 
\citet{2003ApJ...586.1067H} find a fraction of $48$\%. Figure~\ref{f:vrmsunm}
shows the mass fractions for
all models against the velocity dispersion in the CNM, averaged
over a time interval $7.5<t<9.5$~Myr. Higher fractions of thermally unstable
gas are found at higher velocity dispersions, a correlation already pointed out earlier
(e.g. \citealp{2001ApJ...557L.121G,2005A&A...433....1A,2006ApJ...648.1052H,2007A&A...465..431H}). 
However, the magnetic field
introduces a second dependency: of the three models with the highest velocity dispersion
(H, Y05 and XR25), model Y05 has the highest fraction of thermally unstable gas.
As discussed above, in this model the lateral field components reduce the density contrasts,
keeping a substantial fraction of the gas in the thermally unstable regime. The large
velocity dispersion of model Y05 arises from the large-scale NTSI interchange mode
(see Fig.~\ref{f:polmap}). 

It is maybe not surprising that the structures which bear closest resemblance to 
the CNM clouds studied by \citet{2003ApJ...586.1067H,2005ApJ...624..773H}
belong to the model with a thermally unstable mass fraction closest to the observed 
one (Y05). 

\section{Discussion}

\subsection{The Role of  Magnetic Fields for Cloud Formation}

The field strength, and the orientation of the mean magnetic field with respect to the 
flows sweeping up the gas play a crucial role for the flow-driven formation of molecular clouds
(Fig.~\ref{f:polmap}). If the fields are dynamically dominant, the only chance to 
build up substantial clouds is by channeling the flows along the fields. 
This is the situation shown in model X50, and it is also 
borne out by simulations of molecular cloud formation in galactic spiral arms
\citep{2006ApJ...646..213K}, where the clouds tend to be oriented perpendicularly to 
the large-scale field (also possibly visible in the models by 
\citealp{2008MNRAS.383..497D}), until sufficient material has been accumulated that they
decouple dynamically from the large-scale field. Similarly, for the sweep-up
of material by e.g. H{\small{II}}-regions or supernova shells, one would expect the densest clouds 
to appear at the locations where the field is perpendicular to the shell
(see Fig.~2 of \citealp{2006ApJ...641..905H}, although the effect might
be less clear in a highly turbulent environment, see Fig.~10 of 
\citealp{2004ApJ...617..339B}). 

\citet{2001ApJ...562..852H} point out -- based on simulations by
\citet{1995ApJ...455..536P} -- that dynamically weak (but not necessarily
ordered) fields would lead to a general selection effect for the formation of molecular
clouds. Since $\beta_{ram}>1$, the flows stretch out the
fieldlines, leading to a natural alignment. In this picture, clouds form in the bends
of large-scale fields (see Figs. 4 \& 5 of \citealp{2001ApJ...562..852H}).
Such a scenario is to some extent
addressed by model XR25, where  varying field orientations entail 
a local selection effect, picking out the formation sites of molecular clouds 
over e.g. a broad shock front. Note that while the field is dynamically weak 
($\beta_{ram}=8.5$, $\beta_{th}=4.3$) in the initial conditions (and in the inflows) of model XR25, 
$\beta_{th} < 1$ within the cloud (Fig.~\ref{f:prssprof}).

\begin{figure*}
  \includegraphics[width=\textwidth]{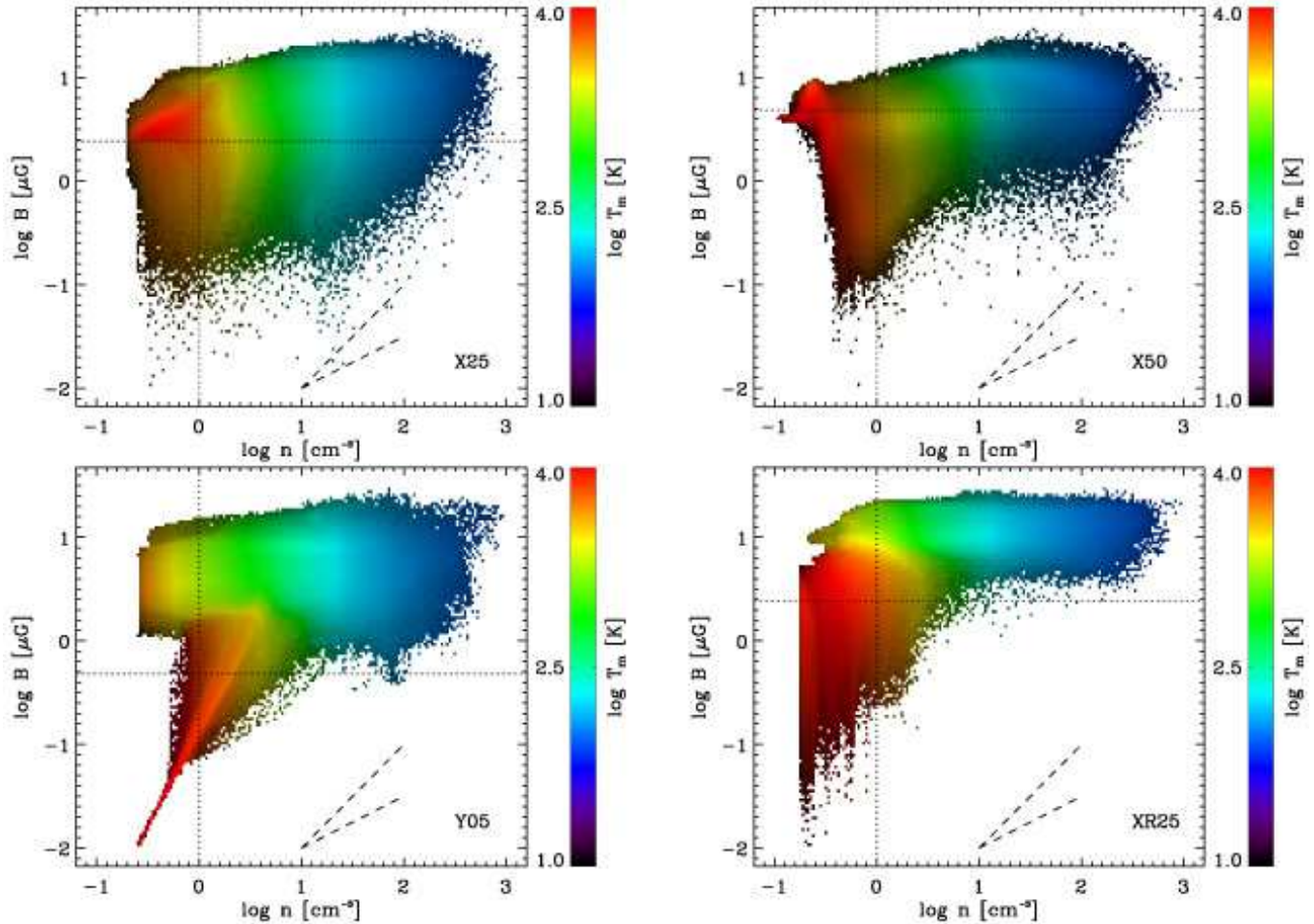}
  \caption{\label{f:babsdens}Magnetic field strength against volume density for models
           as indicated in the plots. The colors denote temperatures, and the intensity
           the mass fraction. Generally, there is no clear correlation between field strength
           and density. The steeper of the two dashed lines denotes $B\propto n$, the 
           flatter one $B\propto n^{1/2}$. Dotted lines denote the initial conditions.}
\end{figure*}

This selection effect comes about because already a small oblique component can
be amplified sufficiently to withstand the compression, preventing the high densities
needed for cloud formation \citep{2001ApJ...562..852H}. Such a situation is addressed in
the extreme by model Y05. 
A field perpendicular to the sweeping-up flow can suppress the formation of massive clouds,
although the three-dimensional situation is much less clear-cut than its one-dimensional
counterpart (see e.g. \citealp{1980ARA&A..18..219M}; \citealp{2004ApJ...612..921B}). 
In one dimension, a density increase by a factor of $100$ from e.g. $n=1$~cm$^{-3}$ to
$100$~cm$^{-3}$ would entail the same factor for the magnetic field strength since $B\propto n$. 
Figure~\ref{f:babsdens} shows this is not the case in three dimensions.
The weak perpendicular field (model Y05) has been amplified by a peak factor of $\approx 30$, 
while the density has increased by up to a factor of $300$. Generally, our models 
show a weak correlation of field strength with density over the whole thermal range, from 
the WNM to the CNM, consistent with observations of the field-density relation 
in the WNM and CNM (\citealp{1986ApJ...301..339T}; HT05), and with
numerical results (e.g. \citealp{2005A&A...436..585D,2008A&A...486L..43H}).
For models X50 and X25 a weak correlation between field and density is not overly surprising.
For model Y05, the decorrelation\footnote{The seemingly strongly correlated B(n) for $n<1$~cm$^{-3}$
in model Y05 does not affect the argument. These are a few regions (low mass fraction) at the edges
of the expanding slab, subjected to numerical reconnection.} is a consequence of the fact that material is still free to 
move along the field lines perpendicular to the original inflow \citep{2007ApJ...665..445H}, 
thus leading to the build-up of 
higher-density filaments perpendicular to the field (but aligned with the inflow), see
Figure~\ref{f:polmap}. Also, other effects, such as the acceleration of magnetic field
transport by turbulence (\citealp{2002ApJ...567..962Z,2002ApJ...570..210F,2004ApJ...603..165H}
for ion-neutral drift, \citealp{1999ApJ...517..700L} for reconnection),
or a decorrelation due to MHD waves \citep{2003A&A...398..845P}
could explain the observed weak correlation.

Magnetic fields will rarely be completely uniform. Model XR25 tests the more
general case of a uniform field at $2.5\mu$G and a random component of equal size, consistent
with (although slightly lower than) observational estimates for magnetic field strengths in the diffuse
gas \citep{1996ASPC...97..457H,2004Ap&SS.289..293B,2006ChJAS...6b.211H}. 
\citet{2007ApJ...663L..41G} 
showed in a two-dimensional numerical experiment that pre-existing  perturbations in the 
inflows can lead to substantial magnetic field amplification due to a rippling of the 
shockfront and subsequent fieldline stretching. We observe a similar effect in model XR25, 
although our Mach numbers are substantially lower (their study addressed the propagation of a 
supernova shock front). \citet{2008A&A...486L..43H} perturb the velocities of the inflows and 
find only a modest increase of the field strength. Clearly, the initially tangled field leads 
to rather different dynamics in the forming cloud (Figs~\ref{f:polmap}, \ref{f:prsstime}). 

Models X25 and X50 demonstrate that not only
the field orientation will play a role during cloud formation (see model Y05), but also
the field strength, since all the instabilities involved have threshold limits for the
field strength -- at least in two dimensions. It might well be that the stronger field
in model X50 suppressing the formation of filaments could be offset by higher inflow speeds
or substructures in the flows. This remains to be studied. 

\subsection{Turbulence and Thermal States}\label{ss:turbtherm}
Fields aligned with the inflows tend to suppress the NTSI, and thus lead to an approximate 
equipartition between the spatial components of the kinetic energy in the cold and
in the thermally unstable gas 
(Fig.~\ref{f:prssprof}, bottom couple of rows). For comparison, the hydrodynamical model 
H has the bulk of the kinetic energy in the (flow-aligned) $x$-component. Magnetic fields may play
an important role to isotropize highly directional flows. Thus, searching for observational
signatures of flow-driven cloud formation should focus on the warm, diffuse gas phase, 
since the inflow signature will be erased in the cold dense gas.

Fractions of thermally unstable gas (Fig.~\ref{f:vrmsunm}) for flow-aligned
fields (models X50, X25) are lower than values observed for diffuse CNM 
clouds \citep{2003ApJ...586.1067H}. A lateral field component results in
a thermally unstable gas fraction of $\approx 50$\%, consistent with observations.
Based on these findings, one could feel tempted to extend the above argument
about the selection effect introduced by magnetic fields: not only could
magnetic fields control the locations of molecular cloud formation, but they
also could lead to ``failed'' molecular clouds, i.e. diffuse atomic hydrogen
clouds, if there is a non-negligible field component perpendicular to the
sweeping-up flow (see also \citet{2008ApJ...687..303I}
for a similar argument based on two-dimensional simulations). 

\subsection{Ordered vs. Random Component}\label{ss:components}

Another observational constraint is given by the ratio of the ordered 
over the unordered (or turbulent) field component. The observational
evidence points to the components being of similar magnitude 
(e.g. \citealp{1996ASPC...97..457H}; \citealp{2004Ap&SS.289..293B};
\citealp{2004ASSL..315..277B}; \citealp{2006ChJAS...6b.211H}; see also
discussion in HT05). A direct comparison to
our models is hampered by the fact that in order to see the varying 
component, sufficiently large scales need to be addressed, which is why
HT05 argue that their observed median field strength of $6\mu$G should
be identified with the {\em total} magnetic field strength. Likewise,
it is not obvious that the components should be of equal magnitude locally 
everywhere. Bearing this limitation in mind, it is clear from 
Figure~\ref{f:fieldtime} that only for models X25 and Y05 the components are 
comparable. 

\subsection{Gravity}

We deliberately left out self-gravity in our simulations, in order to get a clearer 
view of the role of magnetic fields during the early cloud formation phase. Thus, our
clouds are only confined by the ram pressure of the inflows, and at later stages, 
the dynamics of the clouds are probably underestimated since gravity as a source
of turbulence is missing (e.g. \citealp{2008MNRAS.385..181F}). 
As a result of the restricted physics, a comparison of our models with observations
is only meaningful for models where gravity is not expected to play a role, i.e. 
for model Y05 addressing the formation of diffuse HI clouds. For all other models, 
we expect gravity to be relevant during the cloud formation process \citep{2008ApJ...689..290H}.

\section{Summary}

Extending our previous work and complementing a model by 
Hennebelle et al. (\citeyear{2008A&A...486L..43H}; see also \citealp{2008arXiv0808.0986B}),
we study the role of magnetic field strength and orientation on the process
of flow-driven cloud formation. Our models include the usual heating and cooling effects,
allowing rapid fragmentation of the flows, they use uniform inflows to study the most unfavorable
conditions for structure formation, and they envisage the formation of clouds in two head-on
colliding flows, i.e. the extreme case for building up massive clouds. We do not include
self-gravity, focusing on the early stages of cloud formation, during which gravity
might be less important. 

Under these conditions, we find that the effects of magnetic fields on the morphology and
on the thermal state of the resulting clouds depend very strongly not only on the field
orientation with respect to the inflow, but also on the field strength. Initial field energies
are below equipartition with the kinetic energies (by a factor of $4.3$, corresponding
to a field strength of $5\mu$G for our flow parameters) even for the strongest field
case in our study (model X50), yet they result in significantly different cloud properties
than those for a field weaker by a factor of $2$ (model X25, $2.5\mu$G). 
Magnetic fields also lead to a redistribution of the inflow energy to the transverse spatial 
directions (Fig.~\ref{f:prsstime}). 
Hence searching for signatures of colliding flows should focus on the diffuse
gas phase, since the cold gas will have no memory of the original flow direction.

Not surprisingly, weak magnetic fields ($0.5\mu$G) perpendicular to the inflows can suppress the build-up
of massive clouds (model Y05). Yet substructure still can arise in the post shock gas, in 
the form of diffuse filaments perpendicular to the field, and of wave-like patterns 
(possibly magnetosonic waves). 
The filaments are a consequence of lateral gas transport 
(see also \citealp{2007ApJ...665..445H}; \citealp{2008ApJ...687..303I}). 
The straight-forward correlation $B\propto n$ is not obeyed (Fig.~\ref{f:babsdens}).
Mass fractions of thermally unstable gas for the model with a lateral field component (Y05)
are consistent with observed values for diffuse HI clouds \citep{2003ApJ...586.1067H}. For
all other models, the fractions are lower (Fig.~\ref{f:vrmsunm}). The ratio of ordered
vs. random field component is consistent with observations only for the weak-field model
X25, and for the diffuse HI cloud model Y05 (Fig.~\ref{f:fieldtime}).

A weak ($2.5\mu$G) uniform field together with a random component of equal size 
leads to a strong over-pressurization of the cloud due to a combined 
increase of magnetic and kinetic pressure (Fig.~\ref{f:prssprof}), with the magnetic
pressure dominating the thermal pressure within the cloud.
High column densities are assembled at locations
where the perpendicular field component is weakest over time. Thus, a tangled field can lead
to a selection effect for cloud formation while not preventing it globally. 

Our numerical models address the ideal MHD limit, i.e. we do not take into account ion-neutral
decoupling or resistive dissipation. It remains to be seen how non-ideal MHD processes affect
the structure formation during the build-up of the clouds (e.g. \citealp{2008ApJ...687..303I}).

\acknowledgements
We thank the referee for a critical and very helpful report.
Computations were
performed at the National Center for Supercomputing Applications
(AST 060034) and on the local PC cluster Star, perfectly maintained and
administered by J.~Hallum \& R. Bonser. FH is supported by the University
of Michigan and NSF grant AST 0807305.
This work has made use of NASA's Astrophysics Data System.

%%%%%%%%%%%%%%%%%%%%%%%%%%%%%%%%%%%%%%%%%%%%%%%%%%
%
%\references
%
%%%%%%%%%%%%%%%%%%%%%%%%%%%%%%%%%%%%%%%%%%%%%%%%%%

\bibliographystyle{apj}
\bibliography{./references}

\begin{thebibliography}{69}
\expandafter\ifx\csname natexlab\endcsname\relax\def\natexlab#1{#1}\fi

\bibitem[{{Audit} \& {Hennebelle}(2005)}]{2005A&A...433....1A}
{Audit}, E. \& {Hennebelle}, P. 2005, \aap, 433, 1

\bibitem[{{Ballesteros-Paredes} \& {Hartmann}(2007)}]{2007RMxAA..43..123B}
{Ballesteros-Paredes}, J. \& {Hartmann}, L. 2007, Revista Mexicana de
  Astronomia y Astrofisica, 43, 123

\bibitem[{{Ballesteros-Paredes} {et~al.}(1999){Ballesteros-Paredes},
  {Hartmann}, \& {V{\'a}zquez-Semadeni}}]{1999ApJ...527..285B}
{Ballesteros-Paredes}, J., {Hartmann}, L., \& {V{\'a}zquez-Semadeni}, E. 1999,
  \apj, 527, 285

\bibitem[{{Balsara} {et~al.}(2004){Balsara}, {Kim}, {Mac Low}, \&
  {Mathews}}]{2004ApJ...617..339B}
{Balsara}, D.~S., {Kim}, J., {Mac Low}, M.-M., \& {Mathews}, G.~J. 2004, \apj,
  617, 339

\bibitem[{{Banerjee} {et~al.}(2008){Banerjee}, {Vazquez-Semadeni},
  {Hennebelle}, \& {Klessen}}]{2008arXiv0808.0986B}
{Banerjee}, R., {Vazquez-Semadeni}, E., {Hennebelle}, P., \& {Klessen}, R.
  2008, ArXiv e-prints

\bibitem[{{Beck}(2004{\natexlab{a}})}]{2004ASSL..315..277B}
{Beck}, R. 2004{\natexlab{a}}, in Astrophysics and Space Science Library, Vol.
  315, How Does the Galaxy Work?, ed. E.~J. {Alfaro}, E.~{P{\'e}rez}, \&
  J.~{Franco}, 277

\bibitem[{{Beck}(2004{\natexlab{b}})}]{2004Ap&SS.289..293B}
{Beck}, R. 2004{\natexlab{b}}, \apss, 289, 293

\bibitem[{{Bergin} {et~al.}(2004){Bergin}, {Hartmann}, {Raymond}, \&
  {Ballesteros-Paredes}}]{2004ApJ...612..921B}
{Bergin}, E.~A., {Hartmann}, L.~W., {Raymond}, J.~C., \& {Ballesteros-Paredes},
  J. 2004, \apj, 612, 921

\bibitem[{{Burkert} \& {Hartmann}(2004)}]{2004ApJ...616..288B}
{Burkert}, A. \& {Hartmann}, L. 2004, \apj, 616, 288

\bibitem[{{Chandrasekhar}(1961)}]{1961hhs..book.....C}
{Chandrasekhar}, S. 1961, {Hydrodynamic and hydromagnetic stability}
  (International Series of Monographs on Physics, Oxford: Clarendon, 1961)

\bibitem[{{Cho} \& {Lazarian}(2003)}]{2003MNRAS.345..325C}
{Cho}, J. \& {Lazarian}, A. 2003, \mnras, 345, 325

\bibitem[{{de Avillez} \& {Breitschwerdt}(2005)}]{2005A&A...436..585D}
{de Avillez}, M.~A. \& {Breitschwerdt}, D. 2005, \aap, 436, 585

\bibitem[{{Dobbs} \& {Price}(2008)}]{2008MNRAS.383..497D}
{Dobbs}, C.~L. \& {Price}, D.~J. 2008, \mnras, 383, 497

\bibitem[{{Elmegreen}(2000)}]{2000ApJ...530..277E}
{Elmegreen}, B.~G. 2000, \apj, 530, 277

\bibitem[{{Elmegreen}(2007)}]{2007ApJ...668.1064E}
---. 2007, \apj, 668, 1064

\bibitem[{{Engargiola} {et~al.}(2003){Engargiola}, {Plambeck}, {Rosolowsky}, \&
  {Blitz}}]{2003ApJS..149..343E}
{Engargiola}, G., {Plambeck}, R.~L., {Rosolowsky}, E., \& {Blitz}, L. 2003,
  \apjs, 149, 343

\bibitem[{{Evans} \& {Hawley}(1988)}]{1988ApJ...332..659E}
{Evans}, C.~R. \& {Hawley}, J.~F. 1988, \apj, 332, 659

\bibitem[{{Fatuzzo} \& {Adams}(2002)}]{2002ApJ...570..210F}
{Fatuzzo}, M. \& {Adams}, F.~C. 2002, \apj, 570, 210

\bibitem[{{Field}(1965)}]{1965ApJ...142..531F}
{Field}, G.~B. 1965, \apj, 142, 531

\bibitem[{{Field} {et~al.}(2008){Field}, {Blackman}, \&
  {Keto}}]{2008MNRAS.385..181F}
{Field}, G.~B., {Blackman}, E.~G., \& {Keto}, E.~R. 2008, \mnras, 385, 181

\bibitem[{{Gardiner} \& {Stone}(2005)}]{2005JCoPh.205..509G}
{Gardiner}, T.~A. \& {Stone}, J.~M. 2005, Journal of Computational Physics,
  205, 509

\bibitem[{{Gardiner} \& {Stone}(2008)}]{2008JCoPh.227.4123G}
---. 2008, Journal of Computational Physics, 227, 4123

\bibitem[{{Gazol} {et~al.}(2001){Gazol}, {V{\'a}zquez-Semadeni},
  {S{\'a}nchez-Salcedo}, \& {Scalo}}]{2001ApJ...557L.121G}
{Gazol}, A., {V{\'a}zquez-Semadeni}, E., {S{\'a}nchez-Salcedo}, F.~J., \&
  {Scalo}, J. 2001, \apjl, 557, L121

\bibitem[{{Giacalone} \& {Jokipii}(2007)}]{2007ApJ...663L..41G}
{Giacalone}, J. \& {Jokipii}, J.~R. 2007, \apjl, 663, L41

\bibitem[{{Han}(2006)}]{2006ChJAS...6b.211H}
{Han}, J.~L. 2006, Chinese Journal of Astronomy and Astrophysics Supplement, 6,
  020000

\bibitem[{{Hanayama} \& {Tomisaka}(2006)}]{2006ApJ...641..905H}
{Hanayama}, H. \& {Tomisaka}, K. 2006, \apj, 641, 905

\bibitem[{{Hartmann} {et~al.}(2001){Hartmann}, {Ballesteros-Paredes}, \&
  {Bergin}}]{2001ApJ...562..852H}
{Hartmann}, L., {Ballesteros-Paredes}, J., \& {Bergin}, E.~A. 2001, \apj, 562,
  852

\bibitem[{{Heiles}(1996)}]{1996ASPC...97..457H}
{Heiles}, C. 1996, in Astronomical Society of the Pacific Conference Series,
  Vol.~97, Polarimetry of the Interstellar Medium, ed. W.~G. {Roberge} \&
  D.~C.~B. {Whittet}, 457

\bibitem[{{Heiles} \& {Troland}(2003)}]{2003ApJ...586.1067H}
{Heiles}, C. \& {Troland}, T.~H. 2003, \apj, 586, 1067

\bibitem[{{Heiles} \& {Troland}(2005)}]{2005ApJ...624..773H}
---. 2005, \apj, 624, 773

\bibitem[{{Heitsch} {et~al.}(2005){Heitsch}, {Burkert}, {Hartmann}, {Slyz}, \&
  {Devriendt}}]{2005ApJ...633L.113H}
{Heitsch}, F., {Burkert}, A., {Hartmann}, L.~W., {Slyz}, A.~D., \& {Devriendt},
  J.~E.~G. 2005, \apjl, 633, L113

\bibitem[{{Heitsch} \& {Hartmann}(2008)}]{2008ApJ...689..290H}
{Heitsch}, F. \& {Hartmann}, L. 2008, \apj, 689, 290

\bibitem[{{Heitsch} {et~al.}(2008{\natexlab{a}}){Heitsch}, {Hartmann}, \&
  {Burkert}}]{2008ApJ...683..786H}
{Heitsch}, F., {Hartmann}, L.~W., \& {Burkert}, A. 2008{\natexlab{a}}, \apj,
  683, 786

\bibitem[{{Heitsch} {et~al.}(2008{\natexlab{b}}){Heitsch}, {Hartmann}, {Slyz},
  {Devriendt}, \& {Burkert}}]{2008ApJ...674..316H}
{Heitsch}, F., {Hartmann}, L.~W., {Slyz}, A.~D., {Devriendt}, J.~E.~G., \&
  {Burkert}, A. 2008{\natexlab{b}}, \apj, 674, 316

\bibitem[{{Heitsch} {et~al.}(2006){Heitsch}, {Slyz}, {Devriendt}, {Hartmann},
  \& {Burkert}}]{2006ApJ...648.1052H}
{Heitsch}, F., {Slyz}, A.~D., {Devriendt}, J.~E.~G., {Hartmann}, L.~W., \&
  {Burkert}, A. 2006, \apj, 648, 1052

\bibitem[{{Heitsch} {et~al.}(2007){Heitsch}, {Slyz}, {Devriendt}, {Hartmann},
  \& {Burkert}}]{2007ApJ...665..445H}
---. 2007, \apj, 665, 445

\bibitem[{{Heitsch} {et~al.}(2001){Heitsch}, {Zweibel}, {Mac Low}, {Li}, \&
  {Norman}}]{2001ApJ...561..800H}
{Heitsch}, F., {Zweibel}, E.~G., {Mac Low}, M.-M., {Li}, P., \& {Norman}, M.~L.
  2001, \apj, 561, 800

\bibitem[{{Heitsch} {et~al.}(2004){Heitsch}, {Zweibel}, {Slyz}, \&
  {Devriendt}}]{2004ApJ...603..165H}
{Heitsch}, F., {Zweibel}, E.~G., {Slyz}, A.~D., \& {Devriendt}, J.~E.~G. 2004,
  \apj, 603, 165

\bibitem[{{Hennebelle} \& {Audit}(2007)}]{2007A&A...465..431H}
{Hennebelle}, P. \& {Audit}, E. 2007, \aap, 465, 431

\bibitem[{{Hennebelle} {et~al.}(2008){Hennebelle}, {Banerjee},
  {V{\'a}zquez-Semadeni}, {Klessen}, \& {Audit}}]{2008A&A...486L..43H}
{Hennebelle}, P., {Banerjee}, R., {V{\'a}zquez-Semadeni}, E., {Klessen}, R.~S.,
  \& {Audit}, E. 2008, \aap, 486, L43

\bibitem[{{Hennebelle} \& {P{\'e}rault}(1999)}]{1999A&A...351..309H}
{Hennebelle}, P. \& {P{\'e}rault}, M. 1999, \aap, 351, 309

\bibitem[{{Inoue} \& {Inutsuka}(2008)}]{2008ApJ...687..303I}
{Inoue}, T. \& {Inutsuka}, S.-i. 2008, \apj, 687, 303

\bibitem[{{Jenkins} \& {Tripp}(2001)}]{2001ApJS..137..297J}
{Jenkins}, E.~B. \& {Tripp}, T.~M. 2001, \apjs, 137, 297

\bibitem[{{Kim} \& {Ostriker}(2006)}]{2006ApJ...646..213K}
{Kim}, W.-T. \& {Ostriker}, E.~C. 2006, \apj, 646, 213

\bibitem[{{Koyama} \& {Inutsuka}(2000)}]{2000ApJ...532..980K}
{Koyama}, H. \& {Inutsuka}, S.-I. 2000, \apj, 532, 980

\bibitem[{{Kunz}(2008)}]{2008MNRAS.385.1494K}
{Kunz}, M.~W. 2008, \mnras, 385, 1494

\bibitem[{{Lazarian} \& {Vishniac}(1999)}]{1999ApJ...517..700L}
{Lazarian}, A. \& {Vishniac}, E.~T. 1999, \apj, 517, 700

\bibitem[{{Lemaster} \& {Stone}(2008)}]{2008ApJ...682L..97L}
{Lemaster}, M.~N. \& {Stone}, J.~M. 2008, \apjl, 682, L97

\bibitem[{{Mac Low} {et~al.}(1998){Mac Low}, {Klessen}, {Burkert}, \&
  {Smith}}]{1998PhRvL..80.2754M}
{Mac Low}, M.-M., {Klessen}, R.~S., {Burkert}, A., \& {Smith}, M.~D. 1998,
  Physical Review Letters, 80, 2754

\bibitem[{{McKee} \& {Hollenbach}(1980)}]{1980ARA&A..18..219M}
{McKee}, C.~F. \& {Hollenbach}, D.~J. 1980, \araa, 18, 219

\bibitem[{{Mouschovias} {et~al.}(1974){Mouschovias}, {Shu}, \&
  {Woodward}}]{1974A&A....33...73M}
{Mouschovias}, T.~C., {Shu}, F.~H., \& {Woodward}, P.~R. 1974, \aap, 33, 73

\bibitem[{{Nakano} \& {Nakamura}(1978)}]{1978PASJ...30..671N}
{Nakano}, T. \& {Nakamura}, T. 1978, \pasj, 30, 671

\bibitem[{{Palotti} {et~al.}(2008){Palotti}, {Heitsch}, {Zweibel}, \&
  {Huang}}]{2008ApJ...678..234P}
{Palotti}, M.~L., {Heitsch}, F., {Zweibel}, E.~G., \& {Huang}, Y.-M. 2008,
  \apj, 678, 234

\bibitem[{{Parker}(1966)}]{1966ApJ...145..811P}
{Parker}, E.~N. 1966, \apj, 145, 811

\bibitem[{{Parker}(1967)}]{1967ApJ...149..517P}
---. 1967, \apj, 149, 517

\bibitem[{{Passot} \& {V{\'a}zquez-Semadeni}(2003)}]{2003A&A...398..845P}
{Passot}, T. \& {V{\'a}zquez-Semadeni}, E. 2003, \aap, 398, 845

\bibitem[{{Passot} {et~al.}(1995){Passot}, {V\'{a}zquez-Semadeni}, \&
  {Pouquet}}]{1995ApJ...455..536P}
{Passot}, T., {V\'{a}zquez-Semadeni}, E., \& {Pouquet}, A. 1995, \apj, 455, 536

\bibitem[{{Roe}(1981)}]{1981JCoPh..43..357R}
{Roe}, P.~L. 1981, Journal of Computational Physics, 43, 357

\bibitem[{{Shu} {et~al.}(1972){Shu}, {Milione}, {Gebel}, {Yuan}, {Goldsmith},
  \& {Roberts}}]{1972ApJ...173..557S}
{Shu}, F.~H., {Milione}, V., {Gebel}, W., {Yuan}, C., {Goldsmith}, D.~W., \&
  {Roberts}, W.~W. 1972, \apj, 173, 557

\bibitem[{{Stone} \& {Gardiner}(2007)}]{2007ApJ...671.1726S}
{Stone}, J.~M. \& {Gardiner}, T. 2007, \apj, 671, 1726

\bibitem[{{Stone} {et~al.}(2008){Stone}, {Gardiner}, {Teuben}, {Hawley}, \&
  {Simon}}]{2008ApJS..178..137S}
{Stone}, J.~M., {Gardiner}, T.~A., {Teuben}, P., {Hawley}, J.~F., \& {Simon},
  J.~B. 2008, \apjs, 178, 137

\bibitem[{{Stone} {et~al.}(1998){Stone}, {Ostriker}, \&
  {Gammie}}]{1998ApJ...508L..99S}
{Stone}, J.~M., {Ostriker}, E.~C., \& {Gammie}, C.~F. 1998, \apjl, 508, L99

\bibitem[{{Troland}(2005)}]{2005ASPC..343...64T}
{Troland}, T.~H. 2005, in Astronomical Society of the Pacific Conference
  Series, Vol. 343, Astronomical Polarimetry: Current Status and Future
  Directions, ed. A.~{Adamson}, C.~{Aspin}, C.~{Davis}, \& T.~{Fujiyoshi}, 64

\bibitem[{{Troland} \& {Heiles}(1986)}]{1986ApJ...301..339T}
{Troland}, T.~H. \& {Heiles}, C. 1986, \apj, 301, 339

\bibitem[{{V{\'a}zquez-Semadeni} {et~al.}(2007){V{\'a}zquez-Semadeni},
  {G{\'o}mez}, {Jappsen}, {Ballesteros-Paredes}, {Gonz{\'a}lez}, \&
  {Klessen}}]{2007ApJ...657..870V}
{V{\'a}zquez-Semadeni}, E., {G{\'o}mez}, G.~C., {Jappsen}, A.~K.,
  {Ballesteros-Paredes}, J., {Gonz{\'a}lez}, R.~F., \& {Klessen}, R.~S. 2007,
  \apj, 657, 870

\bibitem[{{V{\'a}zquez-Semadeni} {et~al.}(2006){V{\'a}zquez-Semadeni}, {Ryu},
  {Passot}, {Gonz{\'a}lez}, \& {Gazol}}]{2006ApJ...643..245V}
{V{\'a}zquez-Semadeni}, E., {Ryu}, D., {Passot}, T., {Gonz{\'a}lez}, R.~F., \&
  {Gazol}, A. 2006, \apj, 643, 245

\bibitem[{{Vishniac}(1994)}]{1994ApJ...428..186V}
{Vishniac}, E.~T. 1994, \apj, 428, 186

\bibitem[{{Zweibel}(1996)}]{1996ASPC...97..486Z}
{Zweibel}, E.~G. 1996, in Astronomical Society of the Pacific Conference
  Series, Vol.~97, Polarimetry of the Interstellar Medium, ed. W.~G. {Roberge}
  \& D.~C.~B. {Whittet}, 486

\bibitem[{{Zweibel}(2002)}]{2002ApJ...567..962Z}
{Zweibel}, E.~G. 2002, \apj, 567, 962

\end{thebibliography}

%%%%%%%%%%%%%%%%%%%%%%%%%%%%%%%%%%%%%%%%%%%%%%%%%%
%%%%%%%%%%%%%%%%%%%%%%%%%%%%%%%%%%%%%%%%%%%%%%%%%%
\end{document}